\documentclass[twocolumn, times, tighten,twocolappendix]{aastex63}

\usepackage{graphicx,multirow,hyperref,url,color,xspace}
\usepackage{amsmath,amssymb}

\newcommand{\msun}{{\rm M}_{\sun}}

\newcommand{\swift}{{\textit{Neil Gehrels Swift}}\xspace}

\newcommand{\source}{{MAXI J1348--630}\xspace}

\newcommand{\appropto}{\mathrel{\vcenter{
  \offinterlineskip\halign{\hfil$##$\cr
    \propto\cr\noalign{\kern2pt}\sim\cr\noalign{\kern-2pt}}}}}

\defcitealias{Carotenuto21}{C21}
\newcommand{\obs}{{\citetalias{Carotenuto21}}\xspace}
\defcitealias{Carotenuto22}{C22}
\newcommand{\car}{{\citetalias{Carotenuto22}}\xspace}

\begin{document}

\title{No Need for an Extreme Jet Energy in the Black-Hole X-Ray Binary MAXI J1348--630}
\shorttitle{The jet in MAXI J1348--630}

\author[0000-0002-0333-2452]{Andrzej A. Zdziarski}
\affiliation{Nicolaus Copernicus Astronomical Center, Polish Academy of Sciences, Bartycka 18, PL-00-716 Warszawa, Poland; \href{mailto:aaz@camk.edu.pl}{aaz@camk.edu.pl}}
\author[0000-0003-1667-7334]{Marek Sikora}
\affiliation{Nicolaus Copernicus Astronomical Center, Polish Academy of Sciences, Bartycka 18, PL-00-716 Warszawa, Poland; \href{mailto:aaz@camk.edu.pl}{aaz@camk.edu.pl}}
\author[0000-0001-7606-5925]{Micha{\l} Szanecki}
\affiliation{Faculty of Physics and Applied Informatics, {\L}{\'o}d{\'z} University, Pomorska 149/153, PL-90-236 {\L}{\'o}d{\'z}, Poland}
\author[0000-0002-8434-5692]{Markus B{\"o}ttcher}
\affiliation{Centre for Space Research, North-West University, Potchefstroom 2520, South Africa}

\shortauthors{Zdziarski et al.}

\begin{abstract}
We model interaction with the surrounding medium of the main discrete jet ejection in the accreting black-hole binary MAXI J1348--630. The kinetic energy in the ejection of that jet was estimated before to be $>10^{46}$\,erg. That energy requires that the jet power was about two orders of magnitude above the limit corresponding to a magnetically arrested accretion onto a maximally rotating black hole. That large estimate was obtained by considering the initial ballistic jet propagation in a surrounding cavity followed by a sudden deceleration in interstellar medium under the assumption of its standard density of $\sim$1\,cm$^{-3}$. Such densities are likely in the surrounding of this source given its location in the Galactic Plane. Here, we show that the estimate of the kinetic energy can be reduced to realistic values of $\sim\! 10^{44}$\,erg by considering the presence of a transition layer with an exponential density growth separating the cavity and the interstellar medium. In that case, the jet is found to decelerate mostly in the transition layer, in regions with the densities $\ll$1\,cm$^{-3}$, which strongly reduces the energy requirement. Still, the required jet masses are large, ruling out the presence of a significant number of electron-positron pairs.
\end{abstract}

\section{Introduction}
\label{intro}

In this Letter, we model the main discrete jet ejection from the low-mass X-ray binary (LMXB) \source during its outburst in 2019. In a previous work, \citet{Carotenuto22}, hereafter \car, measured the kinetic energy in that ejection as $>10^{46}$\,erg, which, in turn, required the jet power to be above the standard maximum of $P_{\rm j}\sim \dot M_{\rm accr}c^2$ (e.g., \citealt{Davis20}) by about two orders of magnitude. Here, $\dot M_{\rm accr}$ is the mass accretion rate onto the black hole (BH), as inferred from the observed accretion emission. If that estimate were correct, at least some jets in BH LMXBs would carry huge amounts of invisible kinetic energy, requiring, in turn, the true accretion rate to be highly supercritical, and questioning our basic understanding of the accretion process in such systems.
  
\begin{table*}
\caption{Initial evolution of \source.}
\label{evolution}      
\centering            
\begin{tabular}{ccc}
\hline 
$T$ range & Radio state & X-ray state \\
\hline 
9--16 & Core, self-absorbed, $\alpha\sim 0$ & Hard, power law with $\alpha<1$ \\
17--19 & Core, self-absorbed, $\alpha\sim 0$ & Hard-intermediate, power law with $\alpha>1$ \\
20--22 & Core, self-absorbed, $\alpha\sim 0$ & Hard-Intermediate, disk blackbody + power law with $\alpha>1$ \\
23 & Core, the main flare, $F_{1.3\,{\rm GHz}}=486$\,mJy & Soft-intermediate, disk blackbody + power law with $\alpha>1$, type-B QPOs \\
$\sim$25 & The estimated RK1 ejection & Soft-intermediate, disk blackbody + power-law with $\alpha>1$, type-B QPOs \\
24--43 & Core, optically-thin, $\alpha\sim 0.3$--0.5 & Soft-intermediate, disk blackbody + power law with $\alpha>1$, type-B QPOs \\
51--67 & No core, RK1 emission, $\alpha\sim 1$ & Soft-intermediate, disk blackbody + power law with $\alpha>1$ \\
\hline                   
\end{tabular}\\ 
{\it Notes:} $T\equiv {\rm MJD}-58500$, the index $\alpha$ defined by the energy flux of $F_\nu \propto \nu^{-\alpha}$. We adopt here the hard-intermediate to soft-intermediate transition time of $T=22.6$, defined by a sudden decrease of the fractional X-ray variability \citep{Zhang20}. 
\end{table*}

\source was discovered in X-rays by the MAXI detector \citep{Matsuoka09} on board of {\it International Space Station\/} on 2019 January 26 (MJD 58509; \citealt{Yatabe19}). Hereafter, we define $T\equiv {\rm MJD}-58500$; thus, the discovery was on $T=9$. The distance to the source has been estimated as $D=2.2^{+0.6}_{-0.5}$\,kpc based on H{\sc i} absorption \citep{Chauhan21}, and  as $\approx 3.4\pm 0.3$\,kpc based on X-ray detections of a dust-scattering halo \citep{Lamer21}. Hereafter, we use $D=2.2$\,kpc as the default value, but consider dependencies of our results on $D$. 

The radio observations of \source by MeerKAT \citep{Jonas16} and the  Australia Telescope Compact Array (ATCA) are described in detail by \citet{Carotenuto21}, hereafter \obs. The evolution of the source in the radio and X-ray bands until the initial evolution of the main ejection, denoted in \obs as RK1, is summarized in Table \ref{evolution}. Up to $T=21.88$--22.06, the 5.5--21.2\,GHz spectrum was hard, $\alpha\approx -0.15$--0.0, characteristic of partially self-absorbed compact jets \citep{BK79}. Here $\alpha$ is defined by the energy flux of $F_\nu \propto \nu^{-\alpha}$. The 5.5\,GHz fluxes during these observations were increasing from a few mJy to $\sim\! 10^2$\,mJy. A bright radio flare was then seen on $T=23.2$ with the 1.3\,GHz flux of $F_\nu\approx 486$\,mJy, and on $T=23.9$, the 5.5--21.2\,GHz spectrum was first seen to be optically thin, with $\alpha\approx 0.5$. The emission in those observations was coming from the direction consistent with the BH position, hereafter referred as the core. Further core emission was observed until $T=43.1$. On $T=51.1$, the RK1 component, displaced from the core by $\approx\! 2.8''\pm 0.5$, was first detected. It followed a linear trajectory up to the separation of $\approx\! 12''$ on $T=121$. It became then invisible until $T=275$, when it reappeared at the separation of $\approx\! 26.7''$, from which point on it was seen moving much slower, up to $\approx\! 28''$ on $T=387$. The RK1 spectrum was optically thin whenever measured, with $\alpha\sim 1$. The observations show it was the approaching component (\obs). The corresponding receding component was not observed, implying a small viewing angle (\obs). 

Then, \car calculated the kinetic energy of the RK1 ejection based on modelling of its interaction with the interstellar medium (ISM). They used a model based on that for $\gamma$-ray afterglows and applied to microquasar jet ejections by \citet{Wang03}. \car found that the jet first traveled at an approximately constant velocity in a low-density cavity, and then entered a standard ISM. They obtained the kinetic energy of the ejection, $E_0$, as
\begin{equation}
E_0\approx 4.6_{-3.4}^{+20.0}\left(\frac{\phi}{1\degr}\right)^2 \frac{n_{\rm ISM}}{1\,{\rm cm}^{-3}}10^{46}\,{\rm erg},
\label{energy}
\end{equation}
where $\phi$ is the jet opening angle, and $n_{\rm ISM}$ is the ISM density outside the cavity. They argued against the possibility that the ejection entered a hot phase of the ISM, in which case the density would be $\ll 1$\,cm$^{-3}$, because the source lies in the Galactic Plane, where high ISM densities are expected. They also argued that $\phi\sim 1\degr$ is a realistic estimate since whenever similar ejecta were resolved, their opening angles were of that order (e.g., \citealt{Miller-Jones04, Rushton17, Espinasse20}). 

As pointed out by \car, and noted above, the value of the kinetic energy of Equation (\ref{energy}) violates by about two orders of magnitude the basic physical constraint based on the maximum jet power of $P_{\rm j}\sim \dot M_{\rm accr}c^2$ with $\dot M_{\rm accr}$ estimated using the observed luminosity and assuming the accretion efficiency of $\epsilon\sim 0.1$, and an estimate of the ejection duration based on the observations. At face value, this indicates that the true accretion efficiency is $\ll\! 0.1$. On the other hand, a strong argument for the the average of $\langle\epsilon\rangle\sim 0.1$ in accreting BH binaries follows from the study of \citet{Coriat12}. Those authors estimated the average accretion rate using the observed X-ray luminosities for a large sample of those sources. They found that using $\langle\epsilon\rangle=0.1$ gives a good agreement with the predictions of the disk instability model for the X-ray transients \citep{Dubus01}. We also note that \source appears to be a typical BH LMXB based on its track on the hardness-luminosity diagram, see fig.\ 1 in \obs. Also, the light curves of the ejection appear similar to those for such events in other sources. Thus, it appears unlikely that either \source or its RK1 ejection are unique.

Still, the conclusion of \car was that Equation (\ref{energy}) provides a correct estimate of the ejection energetics. In order to support this, they noted that \citet{MR94} estimated the kinetic energy of the bulk motion in the 1994 ejection from GRS 1915+105 to be similarly large, $3\times 10^{46}$\,erg. \citet{MR94} also estimated the minimum jet power for GRS 1915+105 as $\sim\! 10^{41}$\,erg\,s$^{-1}$, similar to the jet power required to supply the very large kinetic energy in our case. However, the estimates of \citet{MR94} were based on the jet Lorentz factor estimated from the proper motion of the twin ejecta, which strongly depends on the distance to the source. \citet{MR94} assumed 12.5\,kpc, while the current distance estimate is $8.6_{-1.6}^{+2.0}$\,kpc \citep{Reid14}. The actual estimates of the energy content and jet power for that event are then rather modest $\sim\! 10^{43}$\,erg and $\sim\! 10^{39}$\,erg\,s$^{-1}$ \citep{Zdziarski14d}. Then \car noted that \citet{Steiner12b} found a deceleration of the ejecta in H1743--322 within its cavity, for which they estimated the kinetic energy as $E_0\sim 10^{45}(n_{\rm cavity}/10^{-2}$\,cm$^{-3}$)\,erg. In this case, the inferred energy satisfies the maximum jet power limit for the cavity density of $n_{\rm cavity}\lesssim 10^{-3}$\,cm$^{-3}$, which appears possible.

On the other hand, \citet{Steiner12} used the same method as \car for twin ejecta from XTE J1550--564. That research also showed the presence of a cavity around the central source and deceleration of the ejecta in the ISM outside the cavity. They found $E_0\approx 5.9_{-2.3}^{+3.6}\times 10^{45}$\,erg (for $n_{\rm ISM}=1$\,cm$^{-3}$ and $\phi=1\degr$), which violates the maximum jet power constraint for that source, similar to the case of \source.

In this Letter, we propose a physically realistic solution allowing us to strongly reduce the energy requirements for a moving ejection. Our model is fitted to the RK1 data of \source. However, it is likely that the same model would also reduce the energy requirements for XTE J1550--564.

\section{The model and results}
\label{model}

\subsection{The model of ejection propagation}
\label{propagation}

We follow the formulation of non-radiative jet propagation through the ISM, originally developed for $\gamma$-ray burst afterglows \citep{Piran99}, in the formulation of \citet{Huang99}, but taking into account a difference between the bulk Lorentz factor, $\Gamma$, and the Lorentz factor of the shock front, $\Gamma_{\rm sh}\geq \Gamma$, following \citet{Wang03}. We employ the exact formalism as used by \car. 

The energy conservation equation reads
\begin{equation}
E_0=(\Gamma-1)M_0 c^2+ s(\Gamma_{\rm sh}^2-1)m_{\rm sw} c^2,
\label{E0}
\end{equation}
where $E_0=(\Gamma_0-1)M_0 c^2$ is the kinetic energy of the single moving jet with the initial rest mass $M_0$ and the initial Lorentz factor $\Gamma_0$, $s\approx 0.73-0.38\beta$, $\beta= (1-\Gamma^{-2})^{1/2}$, and $m_{\rm sw}$ is the mass of the outside medium acquired by the ejection. This adiabatic formalism assumes that the internal energy in particles and magnetic fields is negligible, as well as the radiated energy is negligible. We use $\Gamma_{\rm sh}(\Gamma)$ from \citet{Blandford76}, with the adiabatic index approximated as in \citet{Steiner12}. We also compare our results for that $\Gamma_{\rm sh}$ with those for the case with $\Gamma_{\rm sh}=\Gamma$, as follows from the original derivation by \citet{Huang99}. The latter has a simple solution of $\Gamma(m_{\rm sw})$ as a root of a quadratic equation. The kinematic equations for the approaching component and the separation on the sky, $\Delta$, are
\begin{equation}
\frac{{\rm d}z}{{\rm d}t}=\frac{\beta c}{1-\beta \cos i},\quad
\Delta(t)=\frac{z \sin\theta}{D},
\label{kinematic}
\end{equation}
where $z$ is the distance from the BH (in its frame), $i$ is the jet inclination, and $t$ is the photon arrival time measured by the observer.

As in \car, we assume that the ejection propagates first within a cavity with the density of $\ll$1\,cm$^{-3}$. Such densities along ejection trajectories are common for microquasars \citep{Heinz02}. The ejection then enters a much denser medium, which is likely to be a standard ISM, with the density $n_{\rm ISM}\sim 1$\,cm$^{-3}$. \car, following \citet{Steiner12}, assumed a sharp, step-function like, transition between the cavity density and that of the ISM at a distance $z_{\rm c}$.  This is clearly not physically realistic; there would be always some finite transition region. Here, we instead assume that the density at $z_{\rm c}$ starts to exponentially grow with an e-folding distance, $d_z$, until it reaches $n_{\rm ISM}$ at $z_{\rm ISM}$, 
\begin{equation}
n(z)=\begin{cases}
n_{\rm cavity}, &z\leq z_{\rm c};\\
n_{\rm cavity}{\rm e}^\frac{z-z_{\rm c}}{d_z}, &z_{\rm c}\leq z\leq z_{\rm ISM};\\
n_{\rm ISM}, &z\geq z_{\rm ISM},\label{density}
\end{cases}
\end{equation}
where $z_{\rm ISM}=z_{\rm c}+d_z\ln q$, and $q=n_{\rm ISM}/n_{\rm cavity}$. By integrating $n(z)z^2$ over $z$, we find the mass entrained during the propagation of a conical ejection is
\begin{align}
&m_{\rm sw}(z)=\frac{\pi m_{\rm p} n_{\rm cavity}\phi^2}{3}\times\label{msw} \\
&\begin{cases}
z^3, &z\leq z_{\rm c};\\
z_{\rm c}^3+ 3 d_z \left[{\rm e}^\frac{z-z_{\rm c}}{d_z}\left(2 d_z^2-2 d_z z +z^2\right)\right. &\\
\quad \left. +2 d_z z_{\rm c} -2 d_z^2- z_{\rm c}^2\right], &z_{\rm c}\leq z\leq z_{\rm ISM};\\
z_{\rm c}^3+3 d_z \left[q\left(2 d_z^2-2 d_z z_{\rm ISM} +z_{\rm ISM}^2\right)\right. &\\
\quad \left. +2 d_z z_{\rm c} -2 d_z^2- z_{\rm c}^2\right]+ q(z^3-z_{\rm ISM}^3), &z\geq z_{\rm ISM}.
\nonumber
\end{cases}
\end{align}
The limit of $d_z\rightarrow 0$ corresponds to the scenario considered by \car.

We have programmed the above equations using procedures from \citet{numrec92}; in particular, we used the adaptive stepsize routine {\tt odeint} to integrate ${\rm d}z/{\rm d}t$ of Equation (\ref{kinematic}). We then implemented the solution as a fitting function of {\sc xspec} \citep{Arnaud96}. The model parameters are $\log_{10}(E_0)$, $\phi$, $\Gamma_0$, $\log_{10}(n_{\rm cavity})$, $\log_{10}(n_{\rm ISM})$, $i$, $D$, $z_{\rm c}$, the ejection time, $T_{\rm ej}$ (as in \car), and the new parameter, $d_z$. Following \car, we assume $\phi=1\degr$ (with $E_0\propto \phi^2$) and $n_{\rm ISM}=1$\,cm$^{-3}$. 

\subsection{The fitting results}
\label{results}

\begin{table*}
\caption{The results of the model fitting. In all cases, we assume $D=2.2$\,kpc, $i\leq 35\degr$ (implied by the absence of a receding component), and $n_{\rm cavity}\geq 10^{-5}$\,cm$^{-3}$. In Model 1, we fit the entire data set approximately following the assumptions of \car, i.e., with a sharp boundary between the cavity and the ISM and no constraint on $E_0$. In Model 2, we impose a physical constraint of $\log_{10}(E_0)\leq 44.5$. Model 3 is similar to Model 2 except for the exclusion of three outlier measurements (see text), and we consider it to be our best model. The uncertainties correspond to a 90\% confidence level, i.e., $\Delta\chi^2=+2.71$ \citep{Lampton76}, and `'f' denotes a fixed parameter. Note that $E_0\propto\phi^2$. }
\label{fits}      
\centering            
\begin{tabular}{ccccccccccc}
\hline 
Model & $\log_{10}(E_0)$ & $\Gamma_0$ & $\log_{10}(n_{\rm cavity})$ & $\log_{10}(n_{\rm ISM})$ & $\phi$ & $z_{\rm c}$  & $d_z$& $i$ & $T_{\rm ej}$ & $\chi^2_\nu$ \\
& erg &  & cm$^{-3}$ & cm$^{-3}$ & $\degr$ & pc & $10^{-3}$\,pc & $\degr$ & d  \\
\hline
1 & $46.2_{-1.1}^{+0.6}$ & $1.70^{+0.07}_{-0.12}$ & $-3.7_{-1.3}^{+1.4}$ &0f & 1f & $0.49_{-0.02}^{+0.14}$ & 0f & $34^{+1}_{-12}$ & $18.9^{+6.7}_{-3.8}$ & 28.5/23\\
2 & $44.5_{-0.9}^{+0}$ & $1.71^{+0.06}_{-0.07}$ & $-5.0_{-0.0}^{+0.9}$ &0f & 1f & $0.43_{-0.06}^{+0.29}$ & $8.1^{+9.6}_{-8.1}$ & $35^{+0}_{-15}$ & $19.6^{+3.2}_{-4.1}$ & 29.5/22\\
{\bf 3} & $44.5_{-1.1}^{+0}$ & $1.81^{+0.11}_{-0.09}$ & $-4.8_{-0.2}^{+0.9}$ &0f & 1f & $0.42_{-0.09}^{+0.22}$ & $10.7^{+16.4}_{-10.7}$ & $35^{+0}_{-18}$ & $24.7^{+4.0}_{-4.8}$ & 10.3/19\\
\hline                    
\end{tabular} 
\end{table*}

We fit the set of 29 core separation measurements of RK1 as given in \obs. We first follow the assumption of \car of a step-function boundary between the cavity and the ISM, $d_z=0$. Since we use $\chi^2$ fitting instead of the Markov Chain Monte Carlo method they used, we only impose limits on some parameters instead of using their prior distributions. We find the model parameters to be relatively loosely constrained, and thus we assume a fixed source distance of 2.2\,kpc \citep{Chauhan21}, a choice which is conservative as it (approximately) minimizes $E_0$. A major constraint on the model is that of the viewing angle. Since we see only the approaching ejection, the jet is viewed at a relatively small angle. From fig.\ 11 of \obs, we infer $i\leq 35\degr$ at $D=2.2$\,kpc. We also assume, rather arbitrarily, that $n_{\rm cavity}\geq 10^{-5}$\,cm$^{-3}$, though we consider consequences of lower values of $n_{\rm cavity}$ below. Our results are given as Model 1 in Table \ref{fits}. We obtain results similar to those of \car, with some relatively minor differences attributable to our different fitting method and the detailed assumptions. 

\begin{figure*}[t!]
\centerline{
\includegraphics[width=\textwidth]{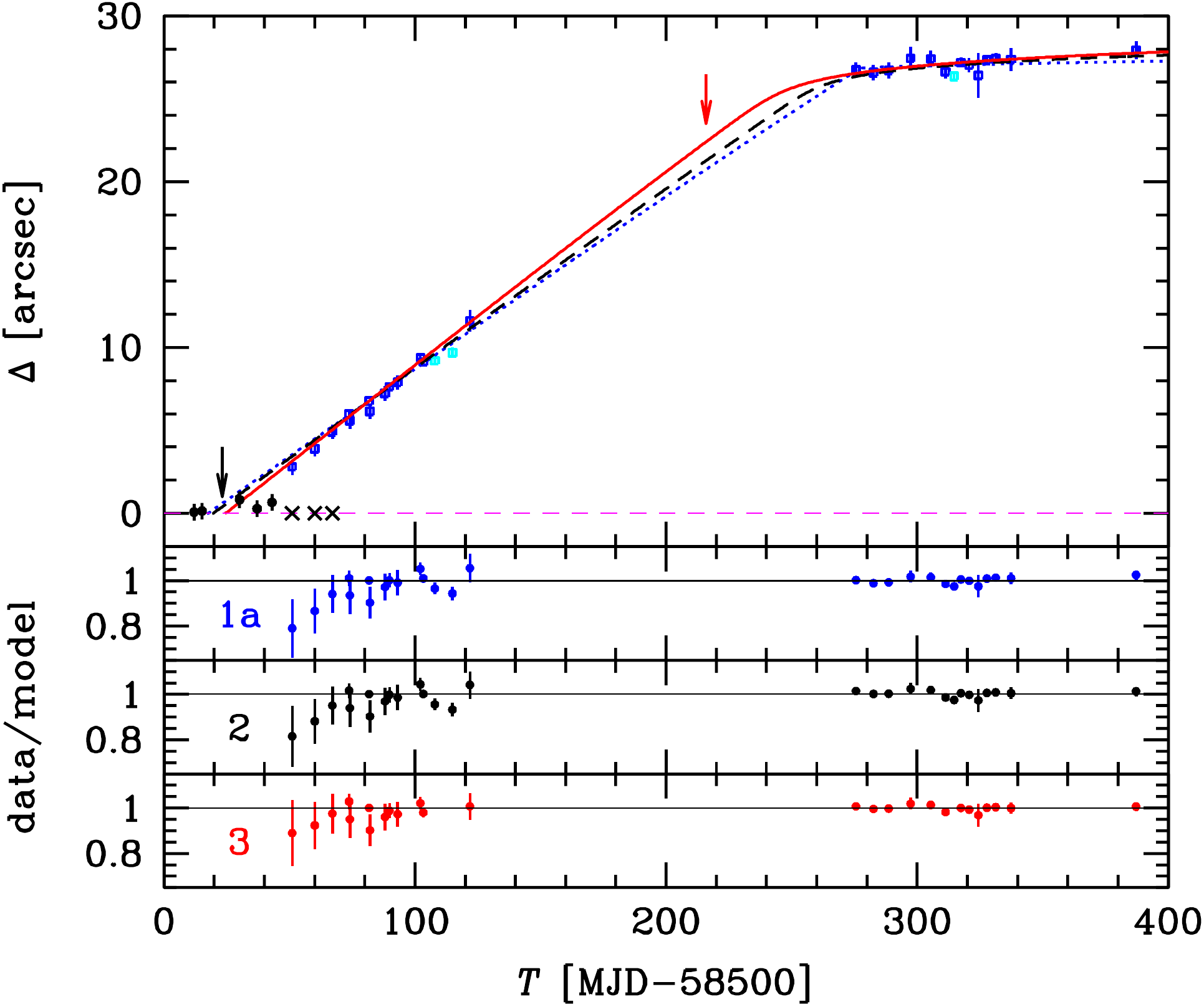}}
  \caption{(Top panel) The observed angular separation, $\Delta$, between the RK1 discrete ejection (blue and cyan squares with error bars; the cyan symbols correspond to the outlier points) and the position of \source as fitted by our models without/with a transition zone and with the physical constraint of $E_0\leq 10^{44.5}$\,erg. The black points with error bars show all of the radio core positions measured prior to the appearance of the radio emission related to the second ejection, RK2, while the black crosses indicate the observations with no core detections; see \obs for details. The black arrow shows the time of the radio flare. The blue dotted curve shows the model fit with a sharp cavity boundary (Model 1a). The black dashed curve shows our fit (Model 2) with $d_z>0$ to the entire data set, and the red solid curve (Model 3, giving the best fit) shows the case excluding the three outlier measurements. In the case excluding the outliers, the initial mass of the jet was $M_0\approx 4.3\times 10^{23}$\,g, and the cavity inner and outer boundaries, $z_{\rm c}\approx 0.41$\,pc and $z_{\rm ISM}\approx 0.54$\,pc, were reached on $T\approx 216$ (marked by the red arrow) and $T\approx 716$, respectively. Thus, the observed reappearance of the jet was entirely within the transition zone. (Bottom panels) The data/model ratios for Models 1a, 2, 3.
}\label{best}
\end{figure*}

In particular, we also obtain a very large ejection energy of $\log_{10}(E_0/{\rm erg})\approx 46.2_{-1.1}^{+0.6}$ (though somewhat lower than the values of \car, see Equation \ref{energy}), at $\chi^2_\nu=28.5/23$. As we discuss in Sections \ref{intro} and \ref{discussion}, we consider such large energies as unphysical. We thus constrain $E_0\leq 10^{44.5}$\,erg, a limit which follows from Section \ref{parameters}, Equation (\ref{energy_d}). In this case, we are still able to obtain a reasonable fit (Model 1a), with $\chi_\nu^2=31.4/23$ ($\Delta\chi^2\approx +2.9$ with respect to the case of unconstrained $E_0$), with $\log_{10}(E_0/{\rm erg})=44.5^{+0}_{-0.6}$. The trajectory of this model are shown in Figure \ref{best} by the dotted blue line, and the residuals are shown by blue symbols. In this model, the deceleration occurs very suddenly when entering the ISM and then the ejection travels with a very small velocity, giving an almost constant angular separation, as seen in Figure \ref{best}. However, as we discuss above, a step-function cavity boundary is not physical. Thus, we allow now $d_z>0$, for which the results are given as Model 2 in Table \ref{fits}, and by the black dashed curve in Figure \ref{best}. This model has $\chi^2\approx 29.5$, close to the original one. 

However, we find that the fits are strongly driven by three outlier measurements, which we define as those each contributing $\chi^2\gtrsim 3$ to the last fit. Those points are on $T\approx 107.9,\, 114.9,\, 314.7$. While including them gives statistically acceptable fits, all three outliers lie below the fitted model. The first two, in particular, cause the initial slope to be shallow and, as a consequence, they significantly reduce the value of $T_{\rm ej}$, to $\approx$20, which is before both the main radio flare and the transition time from optically-thick to optically thin radio spectra, which we consider unlikely. Also, an initial part of the Model 2 trajectory shows systematic data/model residuals, see Figure \ref{best}. 

\begin{figure}
\centerline{
\includegraphics[width=\columnwidth]{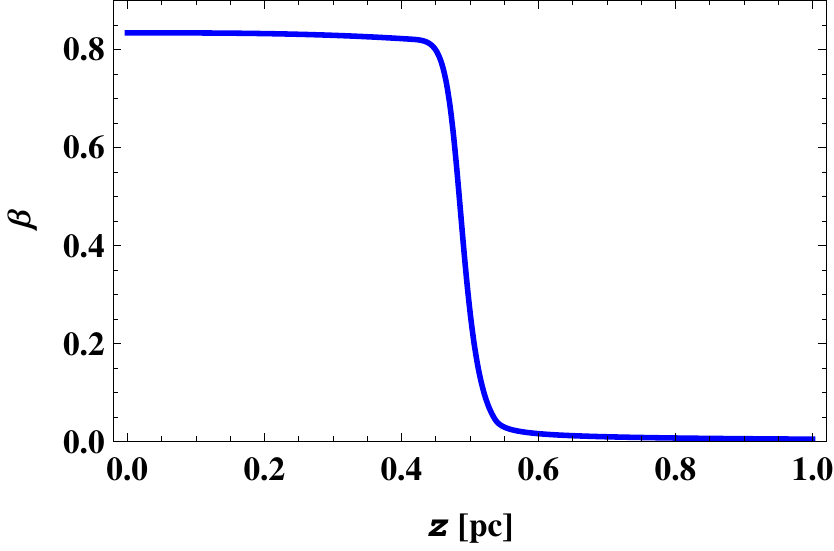}}
  \caption{The dimensionless velocity profile corresponding to our best fit (the red curve in Figure \ref{best} and Model 3 in Table \ref{fits}). The transition region begins at $z_{\rm c}\approx 0.41$\,pc and it ends at $z_{\rm ISM}\approx 0.55$\,pc.
}\label{beta}
\end{figure}

The three outlier points appear to be due to measurement inaccuracies; similar inaccuracies affect the measurements of the core positions. In fig.\ 8 of \obs, there are a number of positions close to the core but with significant displacements, which those authors do not consider to be real (F. Carotenuto, private communication). We thus have performed the fitting without the three outliers. We consider the results without them to reflect the physical reality much better, and we concentrate our further discussion on that case. The results for this fit with the same assumptions as before are given as Model 3 in Table \ref{fits}, and are shown by the red solid curve in Figure \ref{best}. We find that removing those three observational points reduces $\chi^2$ by $\approx$19, to $\chi^2_\nu\approx 10.3/19$. The best-fit ejection energy is still $E_0\approx 10^{44.5}$\,erg. However, if we allow for no upper limit on $E_0$, the $\chi^2$ remains almost unchanged, within $\Delta \chi^2$ of $\pm 0.2$ within $\log_{10}(E_0)\approx 43.7$--46.5. The best fit value of the ejection time is 24.7, which is about a day after the main radio flare and the observed transition to an optically-thin radio spectrum. The velocity profile corresponding to that fit is shown in Figure \ref{beta}.

The physical reason for finding relatively low values of the kinetic energy to be fully compatible with the data as compared to the case of a sharp boundary of the cavity is that the jet now decelerates mostly in an initial region of the exponential density growth, see Figure \ref{beta}. This region still has $n\ll\! n_{\rm ISM}$, which strongly reduces the requirement on the jet kinetic energy (which is proportional to the surrounding density). In our models with the transition zone, the trajectory covered by the observations is entirely within $z<z_{\rm ISM}$. In our last model, the jet mass was doubled over $\approx 5.6 d_z\approx 0.060$\,pc, and the inner and outer boundaries of the cavity, $z_{\rm c}\approx 0.41$\,pc and $z_{\rm ISM}\approx 0.54$\,pc, were reached on $T\approx 216$ and $T\approx 716$, respectively. Thus, the outer boundary was reached a long time after the last observation. In such a case, there is no dependence on $n_{\rm ISM}$, and instead 
\begin{equation}
E_0\propto n_{\rm cavity}\phi^2. 
\label{ncavity}
\end{equation}
Then, for $n_{\rm cavity}=10^{-6}$\,cm$^{-3}$, we obtain $\log_{10}(E_0/{\rm erg})= 43.7^{+0.8}_{-1.0}$. Thus, we can achieve arbitrarily low values of $E_0$ for low enough cavity densities. We note that our specific values of $E_0$ are for the assumption of the exponential density growth of Equation (\ref{density}). While either exponential growth or decline are common in nature, this assumption is uncertain, and it is possible that the density profile in the transition zone is different. This may change the exact value of the fitted $E_0$, and lower it if the transition profile is slower than exponential. Still, the densities first encountered by the jet in that zone will be $\ll n_{\rm ISM}$, which allows for $E_0$ to be low. Also, the proportionality of Equation (\ref{ncavity}) will be preserved. 

We note that we assumed that the jet instantaneously achieved the velocity corresponding to $\Gamma_0$ at the ejection, while it had to be first accelerated from rest. However, we find it is a negligible effect since even $z= 10^7 R_{\rm g}$ (where $R_{\rm g}$ is the gravitational radius), at which $z$ the jet most likely travels with its terminal velocity, corresponds to a light-travel time of $\sim$1000\,s, which is still way below the accuracy of our determination of $T_{\rm ej}$.

We also find that a simpler model with $\Gamma_{\rm sh}=\Gamma$ gives an almost identical dependence of $\Delta(T)$. Thus, our results are very weakly dependent of the details of the treatment of the shock waves excited by the jet. 

\section{Discussion}
\label{discussion}

\subsection{The accretion state}
\label{bol}

Close to the time around the main radio flare ($T=23.2$), the 1--10\,keV unabsorbed fluxes measured by the \swift X-Ray Telescope \citep{Gehrels04} on $T= 23.0$ and 23.45 $\approx$(2.4--$2.6)\times 10^{-7}$\,erg\,cm$^{-2}$\,s$^{-1}$ (\obs). The first and second observation had effective exposure times of 219 and 1412\,s, respectively. We have fitted the spectra in {\sc xspec} using a disk blackbody ({\tt diskbb}; \citealt{Mitsuda84}) model. The disk emission undergoes Compton scattering, leading to a high-energy tail. Following \obs, we modelled the tail as a power law. The spectra are absorbed in the ISM, for which we assumed the column density of $8.6\times 10^{21}$\,cm$^{-2}$, as obtained by \obs using the {\tt tbabs} model \citep{Wilms00}. Thus, the model is {\tt tbabs(diskbb+power)}. We have obtained the inner disk temperatures of $kT_{\rm in}\approx 0.7$--0.8\,keV (similar to \obs), and the normalization of $N_{\tt diskbb}\approx 4.7_{-0.5}^{+0.5}\times 10^4$ and $4.5_{-0.4}^{+0.4}\times 10^4$, for the first and second data set, respectively. In order to estimate the bolometric flux, we used the {\tt thcomp} model \citep{Z20_thcomp}, with a fraction of the disk blackbody being Comptonized. Our model is then {\tt tbabs*thcomp(diskbb)}. Since the fitted range is only 0.7--10\,keV, we cannot constrain the electron temperature, and we have fixed it at either $kT_{\rm e}=50$ or 100\,keV. The unabsorbed bolometric flux (for the 2nd observation, which has much better statistics) is $\approx 4.2\times 10^{-7}$\,erg\,cm$^{-2}$\,s$^{-1}$, insensitive to the choice of $kT_{\rm e}$. The un-scattered flux is dominated by that of the blackbody, which is $\approx 3.5\times 10^{-7}$\,erg\,cm$^{-2}$\,s$^{-1}$. The bolometric flux corresponds to the bolometric luminosity of $2.4(D/2.2\,{\rm kpc})^2 10^{38}$\,erg\,s$^{-1}$. The Eddington luminosity at the cosmic composition (for the hydrogen fraction of 0.7) is $\approx 1.5(M/10\msun)10^{39}$\,erg\,s$^{-1}$, and the Eddington ratio is $\approx 0.16 (D/2.2\,{\rm kpc})^2 (M/10\msun)^{-1}$. The BH mass remains unknown, and we thus scale the results to $10\msun$. 

The inner radius of blackbody-emitting disk in the {\tt diskbb} model is related to $N_{\tt diskbb}$ by
\begin{equation}
R_{\rm in}=x\kappa^2 D N_{\tt diskbb}^{1/2}\cos^{-1/2}\!i,
\label{rin_disk}
\end{equation}
where $x\equiv 10^4{\rm cm}/1\,{\rm kpc}\approx 3.24\times 10^{-18}$, and $\kappa\approx 1.7$ \citep{ST95,Davis19} is the color correction. From that and $N_{\tt diskbb}\approx (4$--$5)\times 10^4$ (as obtained from the {\tt diskbb+powerlaw} fits above), $R_{\rm in}\approx (1.4$--$1.5)(D/2.2\,{\rm kpc}) 10^7$\,cm, or $\approx (9$--$10)(D/2.2\,{\rm kpc})(M/10\msun)^{-1} R_{\rm g}$. Thus, $R_{\rm in}$ is larger than the innermost stable circular orbit for any $a_*$, indicating a disk truncation. This implies a relatively low accretion efficiency \citep{SS73} of $\epsilon\approx R_{\rm g}/(2 R_{\rm in})\sim 0.05$. This gives 
\begin{equation}
\dot M_{\rm accr} c^2\approx 4.8(D/2.2\,{\rm kpc})^2 (\epsilon/0.05)^{-1} 10^{39}\,{\rm erg\,s}^{-1}.
\label{mdotc2}
\end{equation} 

We also consider the appearance of type-B quasi-periodic oscillations (QPOs) in the X-ray light curves around the time of the RK1 ejection. They are often related to ejections of discrete jets \citep{FBG04}. In \source, type-B QPOs were observed during $T\approx 22.6$--40 \citep{Zhang21}. This interval began at the transition between the soft-to-hard intermediate states \citep{Zhang20}, and it closely corresponded to that during which the core radio emission was optically thin, see Table \ref{evolution}. It also included the time of the main radio flare, $T=23.2$, and our estimated ejection time, $T\sim 25$. \citet{Zhang21} found that type-B QPOs were associated with an increase of the flux in the coronal Comptonized component, which is likely to be associated with jet formation. Still, the physics of the connection between type-B QPOs and jet ejection remains uncertain (see also \citealt{Miller-Jones12, Russell19, Wood21} for other cases of this connection).

\subsection{Jet parameters}
\label{parameters}

The maximum possible jet power (for both the jet and counterjet) is achieved for a magnetically arrested flow (MAD; \citealt{BK74, Narayan03, McKinney12}), when it is \citep{Davis20}
\begin{equation}
P_{\rm j}\approx 1.3 h_{0.3} a_*^2 \dot M_{\rm accr} c^2.
\label{pmax}
\end{equation}
Here $h_{0.3}$ is a dimensionless half-thickness of the accretion disk, $h=r\times 0.3 h_{0.3}$ with $h/r\approx 0.3$ being approximately the maximum possible thickness, achievable for hot disks, and $a_*$ is the dimensionless spin parameter. Hereafter, we assume that the jet power is at that maximum. The X-ray spectrum during the ejection was of the intermediate state transiting to the soft state (Section \ref{bol}). While blackbody-emitting disks are typically thin, $h/r\ll 0.3$, we found the disk to be truncated at $\sim\! 10 R_{\rm g}$. The innermost part of the accretion flow is then a hot and radiatively inefficient Comptonizing plasma, which is likely geometrically thick, with $h/r\sim 0.3$. It is also possible that the inner accretion flow consisted of a hot flow with cold, blackbody-emitting, clumps \citep{Liska22}, in which case $h/r\sim 0.3$ is likely as well. We then have
\begin{equation}
P_{\rm j}\approx 6.3 h_{0.3} a_*^2\!\left(\frac{D}{2.2\,{\rm kpc}}\right)^2 \!\left(\frac{\epsilon}{0.05}\right)^{-1}\!\! 10^{39}\,\frac{\rm erg}{\rm s}.
\label{Pj}
\end{equation}

We then consider the mass flow rate through the jets, $\dot M_{\rm j}$. For rest-mass dominated jets, $P_{\rm j}=\dot M_{\rm j}c^2(\Gamma_0-1)$ (where we consider epochs after conversion of most of the initial Poynting flux into acceleration but before the jet deceleration in the surrounding medium). This implies
\begin{equation}
\dot M_{\rm j}=\frac{1.3 h_{0.3} a_*^2\dot M_{\rm accr}}{\Gamma_0-1}.
\label{mdot}
\end{equation}
Note that studies of observed superluminal velocities in blazar radio cores indicate that jets in radio-loud AGNs propagate with the Lorentz factor typically within the range  5--15 (e.g., \citealt{Lister19}). Thus, Equation (\ref{mdot}) challenges the issue of loading a jet by protons at their base in X-ray binaries, where $\Gamma \sim 2$, much more than in radio-loud AGNs \citep{ORiordan18}. It supports the idea that loading of a jet by protons is dominated by its interactions with an MHD wind produced by the innermost portions of the accretion flow \citep{Chatterjee19}.

We stress that $\dot M_{\rm accr}$ in Equation (\ref{pmax}) is the mass flow onto the BH, which is lower than the mass transfer rate from the donor. On the other hand, our estimate of $\dot M_{\rm accr}$ of Equation (\ref{mdotc2}) is based on the disk blackbody emission at radii $\gtrsim\! 10 R_{\rm g}$, which is most likely larger than that on the BH, an effect which reduces $P_{\rm j}$. Given the approximate character of our estimates, that would have a relatively minor effect for $\dot M_{\rm j}\lesssim \dot M_{\rm accr}$. 

We assume the duration of the ejection, $\Delta t_{\rm ej}$, to be given by the time interval between the main flare, $T=23.2$, and the ejection time in our final model, whose best-fit value is $T_{\rm ej}=24.7$. This yields $\Delta t_{\rm ej}\approx 1.5$\,d. However, the fitted $T_{\rm ej}$ has a large uncertainty, and we thus consider $\Delta t_{\rm ej}$ to be a free parameter. We note that this interval corresponds to times measured at the core; thus no frame transformation should be applied to it. Then, $P_{\rm j}/2=E_0/\Delta t_{\rm ej}$, and
\begin{equation}
E_0\approx 4.1 h_{0.3} a_*^2\frac{\Delta t_{\rm ej}}{1.5\,{\rm d}}
\left(\frac{D}{2.2\,{\rm kpc}}\right)^2 \!\!\left(\frac{\epsilon}{0.05}\right)^{-1}\!\! 10^{44}\,{\rm erg}.
\label{energy_d}
\end{equation}
The numerical coefficient above approximately corresponds to the upper limit assumed in Section \ref{results}.

We estimate the length of the jet after the time $\Delta t_{\rm ej}$ as $\beta c \Delta t_{\rm ej}$, which, for $\Delta t_{\rm ej}=1.5$\,d, becomes $\approx 3\times 10^{15}$\,cm. The jet radius at this distance is 
\begin{equation}
R_{\rm j}\approx 6\frac{\Delta t_{\rm ej}}{1.5\,{\rm d}}\frac{\phi}{1\,\degr}10^{13}\,{\rm cm}.
\label{radius_dt}
\end{equation}
As the jet travels ballistically, its radius will increase proportional to the distance, but the jet length could increase only slowly, see Section \ref{radiation}. 

These sizes are well below the limit from the spatial resolution. The observations with the smallest point-spread function (PSF) are those of ATCA at 9\,GHz, where the PSF is $2.5\arcsec$. Taking the source characteristic size as a half of the PSF, we obtain the upper limit of $\lesssim\! 4(D/2.2\,{\rm kpc})10^{16}\,{\rm cm}$. 

The initial mass of the jet in the our fit excluding the outlier measurements is $M_0\approx 4.3\times 10^{23}$\,g, which is approximately $\propto E_0$. We consider the jet composition. If it were dominated by $e^\pm$ pairs, there would be $N_+\approx 2.4\times 10^{50}$ pairs. The required pair production rate is then $\dot N_+\approx 2.8(\Delta t_{\rm ej}/1.5\,{\rm d})^{-1} 10^{45}$\,s$^{-1}$. For comparison, $\dot N_+\sim 2\times 10^{40}$\,s$^{-1}$ was estimated by considering pair production by accretion photons ($\gamma\gamma \rightarrow {\rm e}^+{\rm e}^-$) within the jet base for the luminous hard state of MAXI J1820+070 \citep{Zdziarski22a}, which had a similar bolometric luminosity, but a much harder X-ray spectrum than that of the soft-intermediate state during the ejection, where we found the X-ray power laws with $\alpha\approx 2$. Thus, for the pair dominance we would need to find a mechanism capable of producing pairs at a rate five orders of magnitude higher in spite of the soft observed spectrum. We consider it highly unlikely. On the other hand, in the absence of pairs, we need to find a mechanism of an efficient baryon loading of the jets \citep{ORiordan18}. 

\subsection{An alternative ejection scenario}
\label{alternative}

In Figure \ref{best}, we see that the core emission persisted until $T=43$, which is long after any estimate of $T_{\rm ej}$. At the same time, the emission at the expected RK1 displacement of $\sim\! 2''$ was not detected (\obs). If we take it at face value and discard the possibility that the RK1 emission was very weak at that time, this indicates that the true ejection time was around $T\sim 40$. This could be the case if the discrete jet was initially accelerated to $\Gamma\sim 10$, and later decelerated by interaction by the remnants of the compact, hard-state, jet (as suggested by \citealt{FBG04}). At $\Gamma=10$ and $i=35\degr$, the sky angular velocity is twice as high as that at $\Gamma=1.8$ (Equation \ref{kinematic}), which can account for the observed increase of the angular separation, $\Delta$, from $T=43$ to $T=51$ (at which time RK1 was first detected). 

A large initial $\Gamma$ would reduce the required large value of $\dot M_{\rm j}$, Equation (\ref{mdot}), bringing it in line with radio-loud AGNs. On the other hand, it would increase the energy requirement by an order of magnitude. This is possible for $n_{\rm cavity}\sim 10^{-6}$\,cm$^{-3}$. 

\subsection{Constraints from radiation}
\label{radiation}

\begin{figure}[t!]
\centerline{
\includegraphics[width=\columnwidth]{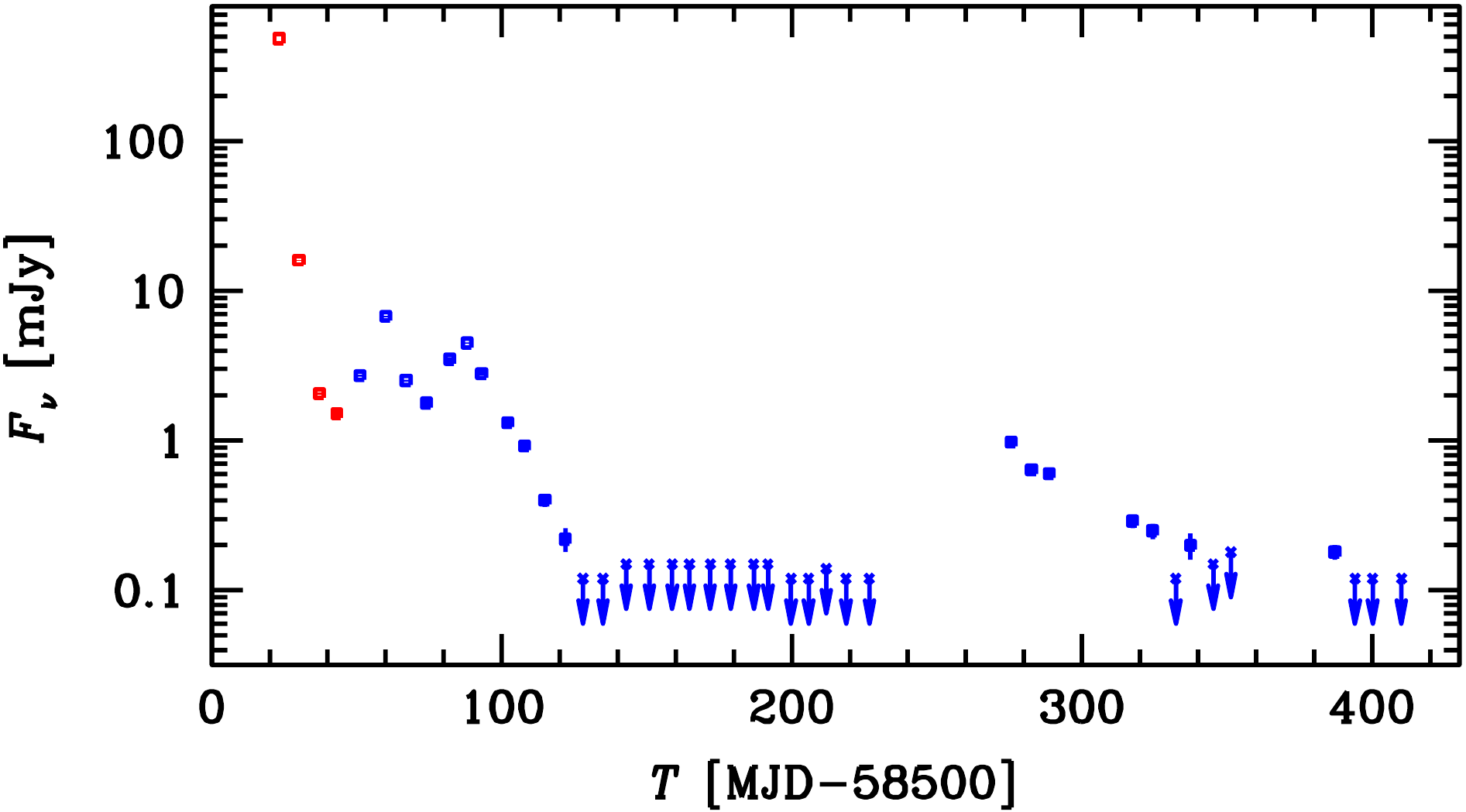}}
  \caption{The fluxes observed at 1.3\,GHz by MeerKAT from the core during the period when ATCA showed optically thin spectra, $\alpha\gtrsim 0.3$ (first four points, red), and from the resolved RK1 (blue points), including the upper limits (blue arrows).
}\label{rk1}
\end{figure}

We consider it likely that the core emission during periods when the spectra were optically thin to come from the same structure as RK1. We show the measured light curve from MeerKAT in Figure \ref{rk1}; we see that the core and RK1 emission join with no more scatter than that in the RK1 light curve alone.

Assuming isotropy in the frame of a moving structure, the comoving spectral luminosity, $L'_{\nu'}$, at $\nu'=\nu/\delta$ is
\begin{equation}
L'_{\nu'}=4\pi D^2 \delta^{-3} F_\nu,
\label{Lnu}
\end{equation}
where $\delta=\left[\Gamma(1-\beta \cos i)\right]^{-1}$ is the Doppler factor. We consider epochs after the jet was accelerated to $\Gamma_0$, but before the deceleration in the surrounding medium. Hereafter we assume $D=2.2$\,kpc, $\Gamma=1.8$, $i=35\degr$.

We first consider a constraint from synchrotron self-absorption. The spectrum changed from partially optically-thick in the 5.5--21.2\,GHz range, $\alpha\sim 0$, to optically thin, $\alpha\approx 0.5$, from $T=22$ to $T=23.9$. At $T=23.9$, the flare with $F_\nu\approx 486$\,mJy was seen at $\nu=1.3$\,GHz (see Figure \ref{rk1}), and we will assume this was a signature of the emission region becoming optically thin. For $L'_{\nu'}$ and the absorption coefficient, $\alpha_{\rm S}$, we use equations (6) and (13), respectively, of \citet{Zdziarski14b}, and consider the optical depth from the jet spine, $\tau=\alpha_{\rm S} R_{\rm j}/(\delta\sin i)$. We assume a conical geometry, with the jet length of $Z_{\rm j}$ in the BH frame, and the comoving volume of $V'=\pi R_{\rm j}^2 Z_{\rm j}\Gamma/3$. By requiring that the self-absorption optical depth is $<$1 and the emission is at the observed level, we can derive a constraint on the jet length, $Z_{\rm j}$, at the time of the source becoming optically thin at $\nu$,
\begin{align}
&Z_{\rm j}^2\gtrsim \frac{8\pi^2 C_2(p) m_{\rm e}^{7/2}c^6 F_\nu D^2 B'^{1/2}} {C_1(p) \phi\Gamma\delta^{3/2} \sin i\, B_{\rm cr}^{5/2}h^{7/2} \nu^{5/2}},
\label{size1}\\
&Z_{\rm j}\gtrsim\label{size2} \\ 
&2 \!\left(\frac{F_\nu}{486\,{\rm mJy}}\right)^{\frac{1}{2}}\!
\left(\frac{B'}{1\,{\rm G}}\right)^{\frac{1}{4}}\!\left(\frac{\nu}{1.3\,{\rm GHz}}\right)^{-\frac{5}{4}}\!\left(\frac{\phi}{1\degr}\right)^{-\frac{1}{2}}\! 10^{13}{\rm cm},
\nonumber
\end{align}
for $\alpha=1$. Here, $B_{\rm cr}={2\pi m_{\rm e}^2 c^3/(e h)}\approx 4.414\times 10^{13}$\,G is the critical magnetic field strength, $m_{\rm e}$ and $e$ are the electron mass and charge, respectively. The angle-averaging coefficients $C_{1,2}(p)\sim 1$ are given by, e.g., equations (8--9) of \citet{Zdziarski22a}, and $p=1+2 \alpha$ is the electron power law index, defined by the electron distribution, $N(\gamma)=K\gamma^{-p}$, where $K$ is the normalization constant. If the jet at that time was stationary, the transformation of Equation (\ref{Lnu}) would be different, and $Z_{\rm j}$ would be higher by $\sim$2.

We then consider later epochs, when the emission became much weaker, with $F_\nu\sim 1$\,mJy. We can estimate the minimum internal energy content in the comoving frame for given optically-thin synchrotron emission by using the method of \citet{Pacholczyk70}, see also \citet{Zdziarski14b}. We assume $\alpha=1$, $\nu=1.3$\,GHz and the power-law emission in the range from $\nu_{\rm min}=0.1$\,GHz to $\nu_{\rm max}=10^3$\,GHz. We obtain
\begin{equation}
E_{\rm e}+E_B\approx 1.7\left(\frac{F_\nu}{1\,{\rm mJy}}\right)^\frac{4}{7} \left(\frac{V'}{10^{45}\,{\rm cm}^3}\right)^\frac{3}{7} 10^{39}\,{\rm erg},
\label{umin}
\end{equation}
where $V'$ is the comoving volume, $E_B/E_{\rm e}=3/4$, and the magnetic field strength is
\begin{equation}
B'\approx 4.3\left(\frac{F_\nu}{1\,{\rm mJy}}\right)^\frac{2}{7} \left(\frac{V'}{10^{45}\,{\rm cm}^3}\right)^{-\frac{2}{7}} \,{\rm mG}.
\label{bmin}
\end{equation}
The above energy content is negligible compared to the rest-mass energy. The electron Lorentz factor dominating the emission at 1.3\,GHz at the default parameters is $\gamma\approx 250$. The minimum particle sound speed is given by
\begin{align}
&\frac{v_{\rm s}}{c}=\sqrt\frac{E_{\rm e}}{3 M_0 c^2}\approx \nonumber\\
&0.0013\left(\frac{F_\nu}{1\,{\rm mJy}}\right)^\frac{2}{7} \left(\frac{V'}{10^{45}\,{\rm cm}^3}\right)^\frac{3}{14}
\left(\frac{M_0}{10^{23}\,{\rm g}}\right)^{-\frac{1}{2}}.
\label{vs0}
\end{align}
This low sound speed would lead to an only moderate jet expansion, e.g., during 100\,d by $\approx 1.0\times 10^{15}$\,cm for the default values above. Note that this estimate neglects the contribution to the particle energy from non-radiating particles. At the default values, the number density of the radiating electrons is $\approx 9\times 10^{-3}$\,cm$^{-3}$, which is $\ll$ that of all electrons, $\approx 110$\,cm$^{-3}$. Thus, a modest contribution to the pressure from the non-radiating particles can increase the estimates above. On the other hand, the jet can be confined by the magnetic field.

We could also calculate the minimum jet power corresponding to the observed emission \citep{Zdziarski14b}. However, as we have found out above, it is likely that most of the jet particles are not radiating, and the jet contains relatively few pairs. Then, the main contribution to the jet power is from the bulk motion of the ions, and the contributions from the magnetic field and internal energy are minor, which makes that method not applicable to RK1.

At the above minimum energy value, the magnetization parameter,
\begin{equation}
\sigma \equiv {B'^2/4\pi\over \rho' c^2},
\label{sigma}
\end{equation}
(where $\rho'$ is the mass density and we assumed the magnetic field is toroidal) is very small, $7\times 10^{-6}$ at the default values. However, $\sigma$ could be larger, with a non-negligible fraction of the jet power being magnetic. In that case, the internal energy in particles would be even smaller than that estimated above. The jet power is the sum of the power in the rest-mass bulk motion, $P_{\rm i}$, and that in the magnetic field, $P_B$. Then, the jet power is given by
\begin{equation}
P_{\rm j}=2\pi R_{\rm j}^2 c \beta\Gamma \left[ \rho' c^2(\Gamma-1) + (B'^2/4\pi)\Gamma\right],
\label{pj}
\end{equation}
which leads to
\begin{equation}
R_{\rm j}=\frac{(P_{\rm j})^{1/2}}{(2 c\beta)^{1/2}\Gamma B'}\left(\frac{\Gamma\sigma}{\Gamma\sigma+\Gamma-1}\right)^{1/2}.
\label{radius}
\end{equation}
Far away  from the origin, most of the initial jet Poynting flu is expected to have converted to the bulk motion, and thus $\sigma< 1$ (e.g., \citealt{Tchekhovskoy09}). Furthermore, $\sigma\ll 1$ is assumed in our kinematic formalism. For $\sigma\lesssim 0.1$, we obtain,
\begin{equation}
R_{\rm j}\approx 1\left(\frac{\sigma}{0.1}\right)^{\frac{1}{2}}\!
\left(\frac{B'}{1\,{\rm mG}}\right)^{-1}\! \left(\frac{P_{\rm j}}{10^{39}\,{\rm erg\,s}^{-1}}\right)^{\frac{1}{2}}\! 10^{17}\,{\rm cm},
\label{radius2}
\end{equation}
which can be constrained by the PSF size limit of $\lesssim\! 4(D/2.2\,{\rm kpc})10^{16}\,{\rm cm}$. On the other hand, assuming an opening angle of $\phi$, the radius at a distance $z$ is
\begin{equation}
R_{\rm j}\approx 1.7\frac{z}{10^{18}\,{\rm cm}} \frac{\phi}{1\degr} 10^{16}\,{\rm cm}.
\label{radius_z}
\end{equation}

\section{Conclusions}

We have found that the radio observations of the RK1 discrete jet from \source do not require unrealistically high kinetic energy. Different from the previous calculation (\car), we consider a transition region between the low-density cavity surrounding the jet and the outside ISM with a relatively high density. For an exponential density growth in that region, we find that the jet loses most of its velocity in an initial part of the transition region, which still has $n\ll n_{\rm ISM}\approx 1$\,cm$^{-3}$. This results in much lower fitted kinetic energies of the jet (proportional to the medium density). 

We estimated the mass accretion rate during the jet ejection based on the X-ray spectrum, which has the form of a disk blackbody with a high-energy tail. Based on that, we estimated the (maximum) jet power corresponding to the MAD accretion, and then the jet energy content based on the ejection duration estimated based on the radio observations and our fit. We have found an agreement with our fitting results provided the cavity density is relatively low, $n\sim 10^{-5}$\,cm$^{-3}$. The likely energy content of the jet is then a few times $10^{44}$\,erg. The corresponding jet mass, $>\! 10^{23}$\,g, is still high, ruling out a substantial content of e$^\pm$ pairs in the jet.

The jet ejection started with the strong radio flare, around which time the radio spectrum became optically thin, indicating an increase of the jet size. After the ejection stopped, the discrete jet was initially narrow, but as it travelled ballistically with an approximately constant opening angle its width increased.

The relatively weak jet emission during the initial part of the trajectory indicates that only a small fraction of electrons is accelerated to relativistic energies. The sound speed of the jet is then very small, allowing the jet expansion to be small (beyond the ballistic motion with a constant opening angle).

Our new method, which takes into account the presence of a cavity--ISM transition layer, can be applied to other observations of discrete ejecta in which they are observed to leave the cavity and enter the ISM. One such case is the BH LMXB XTE J1550--564 \citep{Steiner12}, in which the assumption of a sharp transition also implied a very high kinetic energy of the ejecta. That energy can be reduced if the transition with a finite width is taken into account. See also \citet{Tomsick03, Kaaret03, Corbel05, Yang10, Miller-Jones11, Russell19, Espinasse20} for other cases of decelerating ejecta.

\section*{Acknowledgments}
We thank Francesco Carotenuto for valuable discussions and for providing us with the \swift/XRT data and the positions of the radio core, and the referee for valuable comments. We acknowledge support from the Polish National Science Center under the grant 2019/35/B/ST9/03944. The work of MS is supported by the University of {\L}{\'o}d{\'z} IDUB grant B2211502000094.07. The work of MB is supported by the South African Department of Science and Innovation and the National Research Foundation through the South African Gamma-Ray Astronomy Programme (SA-GAMMA).

\bibliography{../allbib}{}
\bibliographystyle{aasjournal}

\label{lastpage}
\end{document}